\documentclass[aps,prl,twocolumn,superscriptaddress,amsmath,amssymb]{revtex4}
\usepackage{graphicx}

\begin{document}

\title{The Cornerstone Of Spin Statistics Connection: The SU(2)$\times C\times T$ Symmetry}
\author{Biao Lian}
\affiliation{Institute for Advanced Study, Tsinghua University, Beijing, 100084, China}
\date{\today}

\begin{abstract}
We investigate the intrinsic reason for spin statistics connection. It is found that if a free field theory is rotationally (SU(2)) invariant, and has time reversal ($T$) and charge conjugation ($C$) symmetries, it obeys the spin statistics connection, except for a special case. Shr{\"o}dinger equation belongs to this special case, and does not obey spin statistics connection. Further we show that if the energy spectrum of a particle $E(p)$ takes the form of a square root, namely contains branch points in the complex $p$ plane, the particle cannot belong to this special case, and must obey the spin statistics connection. This conclusion includes the relativistic particles as a particular example.

\end{abstract}
\maketitle

The famous spin statistics connection concludes that a particle carrying integral (half-odd integral) spin must have a bosonic (fermionic) statistics. Since it is first proved by Wolfgang Pauli \cite{Pauli}, this theorem has passed various tests and shown to be right. In Pauli's original proof of this theorem, three important assumptions in quantum field theory are used, which are the energy positivity, the Lorentz invariance, and the causality. However, many researchers cast doubts on the necessity of the Lorentz invariance, because it is such a strong condition, and therefore is possibly extraneous for the proof of the spin statistics connection. If one tries to follow any version of the proof of the spin statistics connection, the most difficult and confusing part will be the complex mathematical computations about the Lorentz group. The Lorentz invariance has become the biggest barrier for most people to understand the spin statistics connection.

\begin{figure}[b]
\centering
\includegraphics[width=80mm]{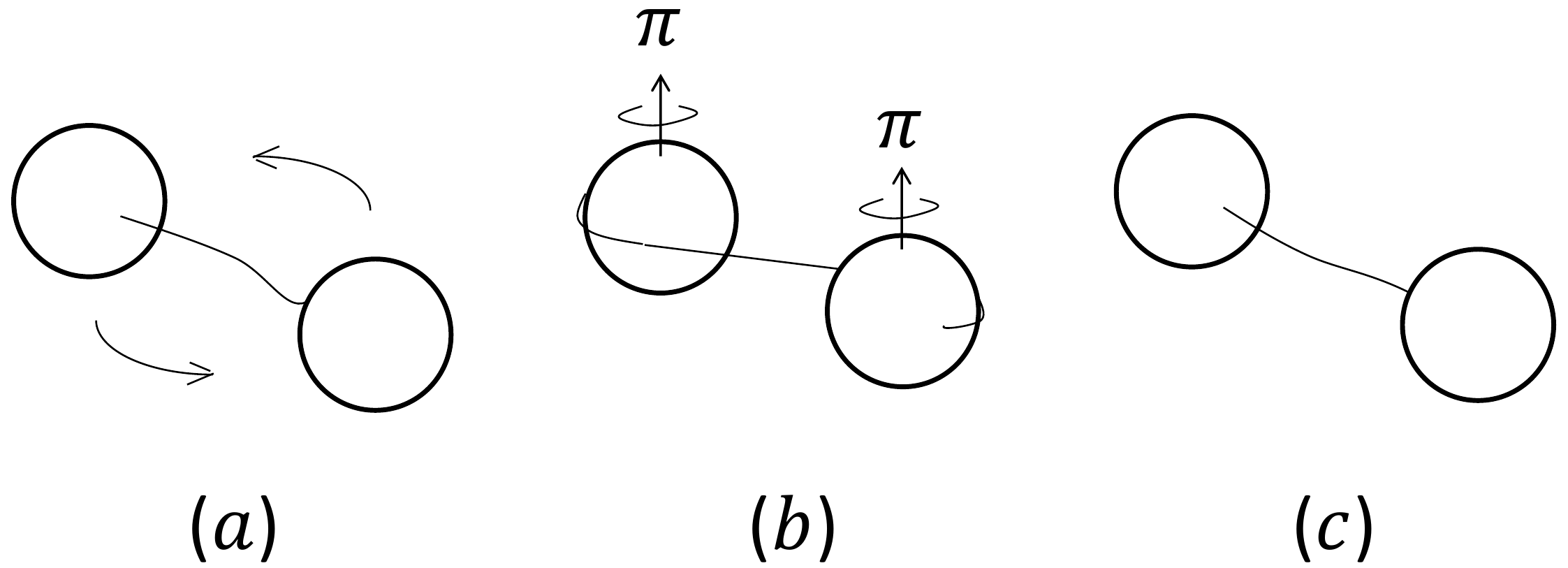}
\caption{A fake picture for the spin statistics connection. Particles are imagined as finite size objects. (a) Imagine there is a line connecting two identical particles. (b) The position exchange does not recover the configuration. (c) One has to continue rotating each particle an angle $\pi$ to recover the initial configuration. The wave function then acquires a phase $e^{-i2\pi j}=(-1)^{2j}$ due to the spin rotation.
\label{fig:F1}}
\end{figure}

On the other hand, many people are more familiar with a ``semi-classical" picture for understanding the spin statistics connection. In this picture, particles are regarded as finite size objects (say, hard spheres) in the space. For description convenience, imagine that there is an invisible line connecting the two identical spin $j$ particles in the way of Figure \ref{fig:F1}(a). When the relative positions of the two particles are exchanged, the original configuration is not yet achieved, as shown in Figure \ref{fig:F1}(b). So one has to continue rotating each of them an angle $\pi$ for their configurations to be recovered (Figure \ref{fig:F1}(c)). Thus their wave function acquires a total phase $e^{-i2\pi j}$ due to the spin rotation, which is $+1\ (-1)$ for integral (half-odd integral) spins respectively. In this incorrect ``proof", the utility of Lorentz invariance is not seen. Naturally people will doubt whether Lorentz invariance is really dispensable for spin statistics connection. Sudarshan ever gives a proof of spin statistics connection for non-relativistic SU(2) invariant free fields \cite{Sudarshan}, and claims that the SU(2) rotation invariance is already enough. This proof has later been doubted and argued by several authors \cite{doubt}. It has also been shown later that under Galilean invariance the relation between spin and statistics of a quantum field is indeed arbitrary \cite{Galilean}. This indicates that spin statistics connection may be violated with only SU(2) invariance. In this paper we investigate the intrinsic reason for spin statistics connection. We will show how spin statistics connection is derived from SU(2)$\times C\times T$ symmetry, which is generally hidden in a free field theory, where $C$ and $T$ stand for charge conjugation and time reversal. We will show the reason of the failure of the spin statistics connection under Galilean invariance, and give the sufficient conditions for the spin statistics connection.

The paper is organized as follows. First we illustrate why the SU(2)$\times C\times T$ symmetry is appropriate for free particles, and construct the general spin $j$ free fields under this symmetry. Then we write down and second quantize the free field Hamiltonian. Next, by requiring the Hamiltonian to be positive definite (vacuum energy equals zero), we show the general derivation of the spin statistics connection, except for a special case that leads to counterexamples. Finally, we analyze the conditions under which counterexamples can arise, and a remarkable corollary of the analysis is, if the energy spectrum $E(p)$ contains branch points in the complex $p$ (momentum) plane, namely, not meromorphic, the spin statistics connection cannot be violated. As a simplest example, a relativistic particle has its energy spectrum $\sqrt{p^2+m_0^2}$ which has branch points at $p=\pm im_0$, so it must obey the spin statistics connection.

\emph{General spin $j$ field equations.} --- The SU(2)$\times C\times T$ invariance is quite a general property of free field equations. SU(2) rotational invariance allows us to define the concept of spin, characterizing different irreducible representations of the SU(2) group. A free field equation also generally has time reversal $T$ invariance, since a free particle motion is always reversible in the absence interactions. In quantum field theory, the time reversal operator $T$ is anti-unitary, and transforms the field operator $\psi$ in the following way
\begin{equation}
T\psi(\mathbf{r},t)T^{-1}=S_T\psi(\mathbf{r},-t) \label{T}
\end{equation}
where $S_T$ is an unitary matrix, depending on the field equation of $\psi$. Charge conjugation $C$ has the meaning of complex conjugation, and has the following form of action on the field operator $\psi$
\begin{equation}
C\psi(\mathbf{r},t)C^{-1}=S_C\psi^*(\mathbf{r},t) \label{C}
\end{equation}
where $S_C$ is also a unitary matrix, while $\psi^*$ is the complex conjugation of $\psi$. One may doubt whether charge conjugation $C$ is a common symmetry of the free field equations, although the answer is ``yes" in relativistic field theories. For the Shr{\"o}dinger equation, which only has the positive energy solutions, it seems hard to define a $C$ symmetry. However, as is observed by Y. Nambu \cite{Nambu}, one could rewrite the Shr{\"o}dinger equation with Nambu spinor $\psi=(\psi^{(S)},\psi^{(S)*})^T$ instead of just the Shr{\"o}dinger spinor $\psi^{(S)}$, as shown in Table \ref{examples}. In this way, one is able to define the $C$ symmetry for this ``doubled Shr{\"o}dinger" equation, whose particles and anti-particles are the same. In general, a free field equation can always be written in a $SU(2)\times C\times T$ invariant form.

Now we wish to construct a general spin $j$ field equation under this symmetry, before we derive the spin statistics connection. Note that a field equation is a group of partial differential equations (PDEs) with spacial and time derivatives. By adding suitable variables, these PDEs can always be reduced to contain no time derivatives higher than the first order (namely, $\partial_t^2$ and higher). For instance, for Klein-Gordon equation $(\partial_\mu\partial^\mu+m_0^2)\phi=0$, we can add $i\partial_t\phi$ as a new variable to rewrite it into two PDEs (see Tab. \ref{examples}). So generally we always can write a field equation in the form below
\begin{equation}
[i\Omega'(\mathbf{p})\partial_t-M'(\mathbf{p})]\psi=0 \label{eq}
\end{equation}
where the field $\psi$ is a column of variables, while $\Omega'(\mathbf{p})$ and $M'(\mathbf{p})$ are matrices which are functions of $\mathbf{p}=-i\nabla$, the momentum operator. For a definite momentum $\mathbf{p}$, a spin $j$ field equation with $C$ and $T$ invariance should have $2j+1$ positive energy solutions and $2j+1$ negative energy solutions. To have exactly these $2(2j+1)$ solutions, $\psi$ should occupy two spin $j$ representations (spin $j$ bispinor), while $\Omega'(\mathbf{p})$ is non-singular in these two representations. Some field equations may have extra variables (components), but they do not represent new freedoms, and must be dependent on this spin $j$ bispinor $\psi$. One can then eliminate them by substitutions when writing the field equation. An example is Proca equation for the four vector $A^\mu$, in which $A^0$ plays the role of an extra variable, as is shown in Tab. \ref{examples}. Therefore we can generally restrict $\psi$ as the composition of two spin $j$ spinors, namely, a spin $j$ bispinor. By defining $\Omega=\Omega'(\mathbf{0})$ and $M(\mathbf{p})=\Omega'(\mathbf{0})[\Omega'(\mathbf{p})]^{-1}M'(\mathbf{p})$, we can rewrite the field equation simpler as
\begin{equation}
[i\Omega\partial_t-M(\mathbf{p})]\psi=0 \label{field-eq}
\end{equation}
where $\Omega$ is now a purely numerical matrix.

We first require Eq. (\ref{field-eq}) to be $SU(2)$ invariant. By Schur's lemma, $\Omega$ must be proportional to the identity matrix in each spin $j$ block. Choosing a proper basis and proper normalization, we can fix the matrix $\Omega$ to be
\begin{equation}
\Omega=\left(\begin{array}{cc}0&I^{(j)}\\I^{(j)}&0\end{array}\right) \label{Omega}
\end{equation}
where $I^{(j)}$ stands for an identity matrix of size $2j+1$. On the other hand, also to ensure the SU(2) invariance, each of the four spin $j$ blocks of $M(\mathbf{p})$ can only be the function of $\mathbf{p}^2$ and $\mathbf{p}\cdot\mathbf{S}^{(j)}$, where $\mathbf{S}^{(j)}=(S_1^{(j)},S_2^{(j)},S_3^{(j)})$ are the spin matrices of spin $j$. The reason is that they are the only invariant scalars under SU(2) rotations.

The $C$ and $T$ symmetries add more limitations to the form of the matrix $M(\mathbf{p})$. For the field Eq. (\ref{field-eq}) to be invariant under $C$ and $T$ transformations, the coefficient matrices must satisfy
\begin{equation}
S_T^\dag\Omega^*S_T=\Omega\ , \qquad S_C^\dag\Omega S_C=-\Omega^* \label{limitA}
\end{equation}
and
\begin{equation}
S_T^\dag M^*(-\mathbf{p})S_T=M(\mathbf{p})\ ,\quad S_C^\dag M(\mathbf{p})S_C=M^*(-\mathbf{p}) \label{limitB}
\end{equation}
For operators $C$ and $T$ to commute with a SU(2) rotation, $S_T$ and $S_C$ must also obey \cite{TimeCharge}
\begin{equation}
S_T^\dag \mathbf{S}^{(j)*}S_T=-\mathbf{S}^{(j)}\ , \qquad S_C^\dag\mathbf{S}^{(j)} S_C=-\mathbf{S}^{(j)*} \label{limitC}
\end{equation}
In addition, the requirement of $C^2=1$ puts another restriction, $S_CS_C^*=I$. With $\Omega$ fixed as in Eq. (\ref{Omega}), limitations (\ref{limitA}) and (\ref{limitC}) are able to determine $S_T$ and $S_C$. Up to a phase factor, one can verify that
\begin{equation}
S_C=\left(\begin{array}{cc}R_2^{(j)}\cos j\pi&R_2^{(j)}\sin j\pi\\-R_2^{(j)}\sin j\pi& -R_2^{(j)}\cos j\pi\end{array}\right) \label{SC}
\end{equation}
while $S_T$ can be fixed to be \cite{Time}
\begin{equation}
S_T=\left(\begin{array}{cc}R_2^{(j)}&\\&R_2^{(j)}\end{array}\right) \label{ST}
\end{equation}
where the matrix $R_2^{(j)}$'s matrix elements are $[R_2^{(j)}]_{m'm}=[e^{-i\pi S_2^{(j)}}]_{m'm}=(-1)^{j+m}\delta_{m',-m}$, and has the property $(R_2^{(j)})^2=(-1)^{2j}$.

\begin{table*}
\caption{\label{examples} Several familiar equations rewritten in the general form shown in Eq. (\ref{field-eq}). The metric for relativistic equations is $g_{\mu\nu}=diag(1,-1,-1,-1)$. The relation between $\psi$ and the old field operator is listed. In the example of Proca equation, $A^0$ is identified as an extra variable. One may check they all have $C$ and $T$ symmetries as shown in Eq. (\ref{SC}) and (\ref{ST}).}
\begin{ruledtabular}
\begin{tabular}{p{13mm}|c|c|c|c}
Name & Klein-Gordon & Dirac & Proca & Shr{\"o}dinger (spin $j$) \\
\colrule
Equation & $(\partial_\mu\partial^\mu+m_0^2)\phi=0$ & $(i\gamma^\mu\partial_\mu-m_0)\psi^{(D)}=0$ & $\partial_\mu(\partial^\mu A^\nu-\partial^\nu A^\mu)+m_0^2A^\nu=0$&$(i\partial_t+\nabla^2/2m_0-\mu)\psi^{(S)}=0$ \\
\colrule
$\psi$ & $(\phi, i\partial_t\phi)^T$ & $\psi^{(D)}$ & $(A^{1},A^{2},A^{3},i\partial_t A^{1},i\partial_t A^{2},i\partial_t A^{3})^T$ & $(\psi^{(S)}+R_2^{(j)}\psi^{(S)*},\psi^{(S)}-R_2^{(j)}\psi^{(S)*})^T$  \\
\colrule
$M(\mathbf{p})$ & $\left(\begin{array}{cc}p^2+m_0^2&\\&1\end{array}\right)$ & $\left(\begin{array}{cc}m_0&\mathbf{p}\cdot\pmb{\sigma}\\-\mathbf{p}\cdot\pmb{\sigma}&m_0\end{array}\right)$ & $\left(\begin{array}{cc}(p^2+m_0^2)I^{(1)}&\\&I^{(1)}\end{array}\right)$ & $\left(\begin{array}{cc}(p^2/2m_0+\mu)I^{(j)}&\\&(p^2/2m_0+\mu)I^{(j)}\end{array}\right)$ \\
\colrule
Extra Variable &  &  & $A^0=i\partial_t(\mathbf{p}\cdot \mathbf{A})/(p^2+m_0^2)$ &  \\
\end{tabular}
\end{ruledtabular}
\end{table*}

Then from limitation (\ref{limitB}), one can easily show that $M(\mathbf{p})$ must take the following form
\begin{equation}
\begin{split}
M&(\mathbf{p})=M_{+}(\mathbf{p}^2,\mathbf{p}\cdot\mathbf{S}^{(j)})\otimes\left(\begin{array}{cc}1&\\ & 1\end{array}\right)\\
&+ M_{-}(\mathbf{p}^2,\mathbf{p}\cdot\mathbf{S}^{(j)})\otimes\left(\begin{array}{cc}\cos j\pi&\sin j\pi \\-\sin j\pi & -\cos j\pi \end{array}\right) \label{M}
\end{split}
\end{equation}

where $M_\pm(x,y)$ are real functions, namely $M_\pm(x,y)=M^*_\pm(x^*,y^*)$. Obviously the two matrices $M_\pm$ commute with each other. By solving Eq. (\ref{field-eq}), one obtain the energy spectrum
\begin{equation}
E^{(\pm)}(\mathbf{p},\sigma)=\pm\sqrt{M_+^2+(-1)^{2j+1}M_-^2} \label{E}
\end{equation}
where $\sigma=\mathbf{p}\cdot\mathbf{S}^{(j)}/p$ is the spin along the momentum $\mathbf{p}$ (or helicity), while $M_\pm$ stands for $M_\pm(p^2,p\sigma)$. We use $u(\mathbf{p},\sigma)$ and $v(\mathbf{p},\sigma)$ to denote the eigenvectors with energies $E>0$ and $E<0$, respectively. These eigenvectors can be easily solved to be
\begin{equation}
u(\mathbf{p},\sigma)=\frac{1}{\mathcal{N}}\left(\begin{array}{c}\sqrt{\sqrt{M_+^2+M_-^2\sin^2 j\pi}-M_-}\ \xi_\sigma \\ \sqrt{\sqrt{M_+^2+M_-^2\sin^2 j\pi}+M_-}\ \xi_\sigma \end{array}\right) \label{u}
\end{equation}
and
\begin{equation}
v(\mathbf{p},\sigma)= \left(\begin{array}{cc}I^{(j)}\cos j\pi&I^{(j)}\sin j\pi \\-I^{(j)}\sin j\pi & -I^{(j)}\cos j\pi \end{array}\right) u(\mathbf{p},\sigma) \label{uv}
\end{equation}
where $\xi_\sigma$ is the spinor satisfying $(\mathbf{p}\cdot\mathbf{S}^{(j)}/p)\xi_\sigma=\sigma\xi_\sigma$, while the normalization factor $\mathcal{N}$ will be determined later. Eq. (\ref{uv}) shows the apparent difference between integral spins and half-odd integral spins.

we have summarized in Table \ref{examples} how several familiar equations are rewritten in the general form above.

Before proceeding to the second quantization procedure, a few more important words on functions $M_\pm$ have to be added. Returning to the original Eq. (\ref{eq}), one should understand both $\Omega'(\mathbf{p})$ and $M'(\mathbf{p})$ as infinite series of $\mathbf{p}$ when really solving the equation. Taking any particular direction $\mathbf{p}=p\mathbf{\hat{n}}$, these two series must be well defined and converge for any $-\infty<p<\infty$. This means both $\Omega'(p)$ and $M'(p)$ cannot contain singularities in the complex $p$ plane. By definition of $M(\mathbf{p})$, we have to require functions $M_\pm(p^2,p\sigma)$ to contain no singularities other than poles in the entire complex $p$ plane, namely, $M_\pm(p^2,p\sigma)$ are meromorphic functions.

\emph{Second quantization and derivation of spin statistics connection.} --- The next step is to second quantize this spin $j$ field $\psi$. The field operator $\psi$ can be expanded according to its eigenvectors as
\begin{equation}
\begin{split}
\psi(\mathbf{r},t)&=\int\frac{\mbox{d}^3\mathbf{p}}{(2\pi)^3}\sum\limits_{\sigma=-j}^{j}\Big[ a_{\mathbf{p},\sigma}u(\mathbf{p},\sigma) e^{i\mathbf{p}\cdot\mathbf{r}-iE(\sigma)t}\\
&+b^\dag_{\mathbf{p},\sigma}v(-\mathbf{p},-\sigma) e^{-i\mathbf{p}\cdot\mathbf{r}+iE(-\sigma)t}\Big]
\end{split} \label{psi}
\end{equation}
where we have used $E(\sigma)$ in short for the positive energy $E^{(+)}(\mathbf{p},\sigma)$, and $a_{\mathbf{p},\sigma}$ and $b_{\mathbf{p},\sigma}$ are the annihilation operators of particles and anti-particles with helicity $\sigma$. For neutral fields with the anti-particle the same as the particle, one should identify these two operators, which does not affect our derivation hereafter. Now we are still unaware whether they obey commutation or anti-commutation relations, namely
\begin{equation}
\frac{[a_{\mathbf{p},\sigma},a^\dag_{\mathbf{p'},\sigma'}]_\pm}{(2\pi)^3}= \frac{[b_{\mathbf{p},\sigma},b^\dag_{\mathbf{p'},\sigma'}]_\pm}{(2\pi)^3}= \delta^3(\mathbf{p}-\mathbf{p'})\delta_{\sigma\sigma'}
\end{equation}
We will show that this is determined by requiring the Hamiltonian to be positive definite and the causality to be satisfied. The Hamiltonian can be derived from the Lagrangian. As one can easily show, the Lagrangian $\mathcal{L}$ for the field Eq. (\ref{field-eq}) is of the form
\begin{equation}
\mathcal{L}=\psi^\dag\Lambda[i\Omega\partial_t-M(-i\nabla)]\psi \label{L}
\end{equation}
where $\Lambda$ is a non-singular, rotationally invariant matrix. For neutral fields with antiparticles identical to particles (such as real Klein Gordon field or Majorana field), $\psi^\dag$ should be replaced by $\psi^T$, which again does not affect the following derivation. By Schur's lemma, each of four blocks of $\Lambda$ is proportional to identity. A physical $\mathcal{L}$ needs to be Hermitian, and this means the three matrices
\begin{equation}
\Lambda\Omega,\ \  M_+\Lambda,\ \
M_-\Lambda\left(\begin{array}{cc}I^{(j)}\cos j\pi&I^{(j)}\sin j\pi \\ -I^{(j)}\sin j\pi & -I^{(j)}\cos j\pi \end{array}\right) \label{la}
\end{equation}
must all be Hermitian matrices. Further, time reversal symmetry leaves us with a requirement $\Lambda^*=\Lambda$.

If $M_-$ is not a zero function (by Eq. (\ref{E}), this also means $|M_+|\ge|M_-|$ is non-zero for integral spin $j$), from the above requirements one can derive out that $\Lambda$ must be
\begin{equation}
\Lambda=\left(\begin{array}{cc}I^{(j)}\cos^2 j\pi&I^{(j)}\sin^2 j\pi \\ I^{(j)}\sin^2 j\pi & I^{(j)}\cos^2 j\pi \end{array}\right)
\end{equation}
With the Lagrangian $\mathcal{L}$, we can then easily write down the Hamiltonian (by substituting Eq. (\ref{psi}))
\begin{widetext}
\begin{equation}
H=\int\mbox{d}^3\mathbf{r}\Big[\partial_t\psi\frac{\partial\mathcal{L}}{\partial\partial_t\psi}-\mathcal{L}\Big] =\int\mbox{d}^3\mathbf{r}\Big[i\psi^\dag\Lambda\Omega\partial_t\psi\Big] =\int\frac{\mbox{d}^3\mathbf{p}}{(2\pi)^3}\sum_{\sigma=-j}^{j} \frac{2E(\sigma)^2}{|\mathcal{N}|^2}[a^\dag_{\mathbf{p},\sigma}a_{\mathbf{p},\sigma}+ (-1)^{2j+1}b_{\mathbf{p},\sigma}b^\dag_{\mathbf{p},\sigma}]
\end{equation}
\end{widetext}
From this result, one can first determine that the normalization factor in Eq. (\ref{u}) is $\mathcal{N}=\sqrt{2E(\sigma)}$. Next, it is clear that to ensure a non-negative $H$, $a_{\mathbf{p},\sigma}$ and $b_{\mathbf{p},\sigma}$ must be commutative (anti-commutative) operators for integral (half-odd integral) spins respectively. One can also immediately demonstrate, with the correct spin statistics relationship, causality is satisfied
\begin{equation}
[\psi(\mathbf{r},t),\psi^\dag(\mathbf{r}',t)\Lambda\Omega]_\pm= \left(\begin{array}{cc}I^{(j)}&\\&I^{(j)} \end{array}\right)\delta^3(\mathbf{r}-\mathbf{r}')
\end{equation}
while it is not with the wrong spin statistics relationship. Till now, we have shown that \emph{spin statistics connection} can naturally result from the SU(2)$\times C\times T$ symmetry hidden in free field theories.

\emph{Counterexamples.} --- However, counterexamples which deviate from spin statistics connection do exist. This is because in the derivation above, we have assumed that $M_-$ is not a zero function. If in an equation $M_-$ is exactly a zero function (but not $M_+$, or the particle will loss its dynamics), condition (\ref{la}) will not be enough to determine the matrix $\Lambda$ uniquely. What we can determine is that $\Lambda$ must be one of the following two matrices
\begin{equation}
\Lambda^{(b)}=\left(\begin{array}{cc}I^{(j)}&\\&I^{(j)} \end{array}\right)\quad,\qquad \Lambda^{(f)}=\left(\begin{array}{cc}&I^{(j)}\\I^{(j)}& \end{array}\right)
\end{equation}
This leads to two possible Lagrangian forms $\mathcal{L}^{(b)}$ and $\mathcal{L}^{(f)}$ correspondingly, which require bosonic and fermionic statistics respectively. As is also easily checked, with $M_-=0$, causality of either commutative or anti-commutative brackets is ensured. Thus for $M_-$ constantly zero, the connection between spin and statistics is arbitrary. As we see in Tab. \ref{examples}, the Shr{\"o}dinger equation is such a counterexample with $M_-=0$, which agrees with the conclusion in \cite{Galilean}.

\emph{Discussions} --- Assume we know the energy spectrum $E(p)$ of a free particle. An immediate corollary is that, if $E(p)$ takes the form of a square root, namely has branch points in the complex $p$ plane, the particle must obey the spin statistics connection. The reason is as follows. As we stated before, functions $M_\pm(p^2,p\sigma)$ do not contain singularities other than poles in the whole complex $p$ plane (meromorphic). Then according to the energy expression (\ref{E}), both $M_+$ and $M_-$ cannot be zero functions, or the energy spectrum can not have any branch points. This corollary naturally applies for relativistic particles whose energy spectrum is $\sqrt{p^2+m_0^2}$, supporting W. Pauli's conclusion \cite{Pauli}.

Though the $M_-$ matrix in Shr{\"o}dinger equation vanishes, in some particular systems it may arise due to interesting mechanisms. For instance, in the mean field theory of superfluid (spin $0$) and superconductivity (spin $1/2$), the rotational invariant order parameter enters into Shr{\"o}dinger equation for the quasi-particles exactly as a $M_-$ matrix.
%In some systems two spin $j$ particles may have tendency of forming a molecule carrying spin $0$, and such a interaction is described by an $SU(2)$ invariant term $g[\Psi^\dag\sum_m\langle0,0|j,m;j,-m\rangle\psi^{(S)}_m\psi^{(S)}_{-m}+h.c.]$. The expectation of the molecule operator $\Psi$ will also play the role of a $M_-$ matrix.
According to our conclusion, these particles or quasi-particles will also have the spin statistics connection.

%Things may become different if we also take into account interactions. In a condensed matter system, quartic order interactions (of order $\psi^4$) may lead to various phases with spontaneous symmetry breaking, such as superfluid for bosons, and superconductivity for fermions. In the mean field treatment of quasi-particles of these phases, the order parameters serve as a non-zero $M_-$ matrix added to the Shr{\"o}dinger equation. This means the quasi-particles must obey spin statistics connection.

%This seems to indicate that a Shr{\"o}dinger equation with interactions may obey spin statistics connection, as interactions looks like an effective $M_-$ matrix. Unfortunately, no one has ever given any proof of spin statistics connection in the presence of interactions, since it is very hard to deal with these interactions.

%Though we have assumed in this paper the particle carries a definite spin $j$, a more general $SU(2)\times C\times T$ invariant field equation need not have a definite spin. On the contrary, a relativistic particle has definite spin because $\mathbf{S}^2$ serves as the Casmir operator of Poincar{\'e} group. But one thing clear is that a free field equation will never mix integral spins with half-odd integral spins, since the spin along momentum $\sigma=\mathbf{p}\cdot\mathbf{S}/p$ is still a conserved quantity.

{\it Acknowledgements.} This work is supported by Tsinghua XueTang Talents Program from the Ministry of Education at Tsinghua University.


\begin{thebibliography}{99}
\bibitem{TimeCharge}
Due to the anti-unitary property of time reversal operator, a unitary transformation $U$ transforms $S_T$ to $U^\star S_T U^\dag$. Under the same unitary transformation $U$, since $S_CU^\star\psi^\star=CU\psi C^{-1}=US_C\psi$, $S_C$ will transform as $U S_C U^T$.
\bibitem{Time}
Under the restrictions, the form of $S_T$ can only be $R_2^{(j)}\otimes e^{i\theta\sigma_x}$, where $\sigma_x$ is the Pauli matrix. By an unitary transformation $U=I^{(j)}\otimes e^{i(\theta/2)\sigma_x}$, both $\Omega$ and $S_C$ is unchanged, while $S_T$ is transformed to Eq. (\ref{ST}).
\end{thebibliography}
\end{document}